\documentclass[twocolumn,showpacs,prb]{revtex4}
\usepackage{graphicx}
\usepackage{dcolumn}
\usepackage{bm}
\def\k{\mbox{\boldmath${\bf k}$}}
\def\G{{\bf \hat G}}
\def\ep{{\bf e }_{spp}}
\def\r{{\Large \mbox{\boldmath$\tau $}}}
\def\u{{\bf u}}

\def\g2{g}
\def\ketbra+-r{ r_p(k_x)\vert{k,p} \rangle^+ \langle{k,p} \vert^- +r_s(k_x) \vert{k,s}\rangle \langle{k,s}\vert}

\def\r{{\bf r} }
\def\p{{\bf p} }
\def\G{{\bf \hat G}}
\def\E{{\bf E} }
\def\sa{\sin\alpha}

\def\Upsilo{ \Phi }

\def\sca{\Theta_{rad}}
\def\scap{\Theta}

\begin{document}
\title{Comparative study of surface plasmon scattering by shallow ridges and grooves }
\author{Giovanni Brucoli}
\email{gianni@unizar.es}
\author{L. Mart\'{\i}n-Moreno}
 \affiliation{ Instituto de Ciencia de Materiales de Arag\'{o}n
 and Departamento de F\'{i}sica de la Materia Condensada,
CSIC-Universidad de Zaragoza, E-50009, Zaragoza, Spain }

\begin{abstract}
We revisit the scattering of surface plasmons by shallow surface
defects for
both protrusions and indentations of various
lengths, which are deemed infinite in one-dimension parallel to the surface.
 Subwavelength protrusions and indentations of equal shape
present different scattering coefficients when their height and
width are comparable. In this case, a protrusion scatters plasmons
like a vertical point-dipole on a plane, while an indentation
scatters like a horizontal point-dipole on a plane. We corroborate
that long and shallow asymmetrically-shaped surface defects have
very similar scattering, as already found with approximate methods.
In the transition from short shallow scatterers to long shallow
scatterers the radiation can be understood in terms of interference
between a vertical and a horizontal dipole. The results attained
numerically are exact and accounted for with analytical models.
\end{abstract}

\maketitle

\section{Introduction}

 Surface Plasmon Polaritons (SPP) are electromagnetic bound modes responsible for
the transport of light at the interface separating a metal from a
dielectric. Their ability to confine light at an air-dielectric
interface offers the prospect of developing a new technology
consisting of photonic
nano-devices\cite{Zia,SPcircuitry,Plasmonic,EkmelOzbay01132006}.
Active research is currently focusing on the possibility of
achieving control over the propagation of SPPs by means of optical
elements that would couple or decouple light to
them\cite{Krenn,Weeber,MUGonzalez,Ilya1,Ilya2}. In order to conceive
optical elements (lenses, mirrors, beam-splitters) able to
manipulate SPPs propagation, we need to learn more about the
interaction of surface plasmons with a sub-wavelength modification
of the underling dielectric metal interface. Indeed the interaction
of SPPs with surface sub-wavelength defects on a metal surface is of
great interest from
a theoretical standpoint.\cite{Zayats,NanoOptics}\\
  In this
article we shall study scattering of SPPs by a \textit{shallow}
surface defect. We will consider both indentations of the metal
surface (grooves) and protrusions on it (ridges). We shall only deal
with bi-dimensional defects, which are deemed infinite in one
dimension parallel to the interface (the
$y$-direction).   
Different aspects of this problem have been studied before with a
variety of numerical techniques
\cite{Pincemin94,FMoreno,PhysRevLett.78.4269,Sanchez-Gil98,SGMaradudin,Sanchez-Gil(OSA),
Sanchez-Gil(APL),
 Ioannis,LevDipole,FLT2005}. Here we present a
systematic comparison between the different scattering coefficients
and provide both analytical expressions and qualitative
explanations.\\ It must be noted that in a previous work we
presented such a comparison\cite{Alexey} but within an approximate
numerical scheme. Within that framework it was found that ridges and
grooves exhibited the same scattering, whenever they are shallow
enough. Here we will revise that result, which turns out to be valid
only for long (elongated) defects. The mistaken outcome of
Ref.~[\onlinecite{Alexey}] for short defects may be traced back to
the breakdown of the assumption of small curvature in the defect
geometry that was made there. In this paper we solve the Maxwell
equations through a discretization method, which does not assume the
previous approximation and whose accuracy depends only on the
discretization mesh. We found that, as in the previous work, long
asymmetric ridges or grooves with the width much larger than the
depth, do scatter very similarly. However square shallow defects
manifest a different scattering efficiency, differing in the
relative radiative loss and radiation pattern. The lack of
distinction between these two cases did not emerge in the previous
approximate treatment. On the whole the problem needs to be
revisited so as to: $i)$ substantiate why the approximate result
does work in the case of elongated defects, $ii)$ point out what is
the correct result in the case of shallow and short symmetric
defects, and $iii)$ explain qualitatively how the scattering
properties of short and shallow symmetric defects are gradually
transformed into the scattering properties of
elongated defects, as the aspect ratio of the defect increases.\\
This paper is organized as follows. In Sec.~\ref{S1} we state the
basic assumptions on the scattering system as well as the solution
method. In Sec.~\ref{Coeff} we rearrange the asymptotic expansions
of the far-field to produce the scattering coefficients. Namely we
express the far-field and the related Poynting vector in terms of
the field inside the defect. Still in this section we look at an
approximation for the scattering coefficients of shallow ridges. In
Sec.~\ref{Ray} we explain that, in general, we cannot quantitatively
represent a scatterer (however small) by one mesh. We explain how we
associate a small symmetric ridge or groove to a point dipole. In
Sec.~\ref{Num} we look at exact numerical results for the scattering
of shallow defects of various horizontal lengths.  We analyze these
results and, in the case of square defects, we associate a ridge to
a vertical dipole and a groove to a horizontal dipole.  In
Sec.~\ref{HV} we produce an analytical model that explains the
radiation pattern of the surface plasmons scattered by small square
ridges and grooves. In Sec.~\ref{NeedleSection} we look at the
solutions for the case of shallow and long defects and we present a
clear-cut interpretation to support the results of the previous
treatment\cite{Alexey}. Finally in Sec.~\ref{Transition} we explain
qualitatively that the aspect ratio of the defect determines the
orientation of the field induced in a shallow defect.
\section{\label{S1}The scattering systems considered}
\begin{figure}
\begin{center}
\includegraphics[scale=1,width=20pc,height=20pc]{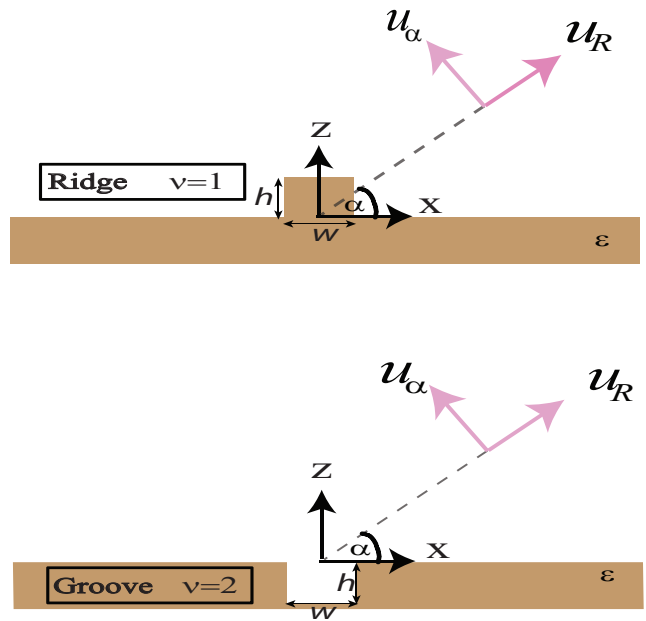}
\caption{Schematic representation of the scattering systems
considered. A ridge is a one dimensional defect located in air and
labeled by the index $ \nu=1$. A groove is a one dimensional defect
located in the metal and labeled by the index $ \nu=2$ }
\label{Scheme}
\end{center}
\end{figure}

The considered defects are infinite in the $y$-dimension and shallow
with depth $h<<\lambda $, where $\lambda $ is the free space
wavelength. The defects are going to be illuminated by a
monochromatic surface plasmon at normal incidence ${\ep}$,
associated to an impinging energy flux $S_{spp}$, defined and
derived in Appendix~\ref{SP}. Therefore, only radiation into
p-polarized(TM) waves needs to be considered.   After we drop, out
of symmetry, the $y$-dependence on the whole problem the field is
expressed as: $\E(\r,t)=\E(x,z)e^{-i \omega t}$.  The wavevector in
vacuum is: $g={2 \pi}/{\lambda}\;$, where  $\omega= c g $. The
material making the slab shall be lossless silver\cite{undici}, that
is: $\varepsilon=\Re\{\varepsilon_{Ag}(\lambda)\}$. Absorption is
neglected as we consider non-resonant defects with widths much
smaller than the SPP propagation length.\\
 As represented in Fig.~\ref{Scheme}, we shall
be expressing the source orientation in a cartesian basis
$(\u_x,\u_z)$, and the scattered fields in a right-handed orthogonal
polar basis: \begin{eqnarray}\label{polar}
\u_{R}&=&\cos\alpha \u_x+\sin\alpha\u_z\\
\u_{\alpha}&=&-\sin\alpha \u_x+\cos\alpha\u_z  \end{eqnarray}
Finally a question of notation: throughout we shall refer to a
bi-dimensional point-source simply as a dipole, but it is meant that
their emission as all of the fields are cyclical in the
$y$-direction. As represented in Fig.~\ref{Scheme}, each object
lying in the vacuum semi-space shall be labeled by the superscript
$\nu=1$ while any object lying in the metal shall be labeled by the
superscript $\nu=2$. In particular, scattering quantities related to
ridges have the superscript $\nu=1$ while the ones related to
grooves have the superscript $\nu=2$.  The field within the
cross-sectional area of the ridge is labeled $\E^r(\r')$ and the one
within that of the groove is labeled $\E^g(\r')$ .

\section{\label{Coeff}Scattering Coefficients }
The Green tensor approach is a standard method to solve
electromagnetic scattering problems
\cite{Hohmann,Protasio,Keller,Li,MGD,Felsen,NanoOptics,AlexImpedance}.
Our first task in this section, is to arrive at an explicit
expression for the scattered electric far-field. This is attained by
propagating the field induced by a dipole density ${\bf
P}^{(1)}(\r')=\Delta\varepsilon\; \E^r(\r')$ (where
$\Delta\varepsilon=\varepsilon-1$) inside the area of a ridge, to a
point ${\bf R}$ very far from the source. For a groove we have the
same relation between polarization and field (except for a change of
sign) ${\bf P}^{(2)}(\r')=-\Delta\varepsilon\; \E^g(\r')$. To
propagate the field from any of the two, we use the standard formula
\cite{NanoOptics}:
\begin{eqnarray}\label{Es}\E_s({\bf R})=g^2
\int_A d\r'\G({\bf R},\r')\cdot{\bf P}(\r').\end{eqnarray}\\
Where $\G({\bf R},\r')$ is the Green tensor for the air-metal
background. The Green tensor propagates the emission of a point
source at $\r' $ to the distant point of the ${\bf R}$. One of the
advantages of the Green tensor technique is that once the fields
inside the defects $\E(\r')$ (and thus ${\bf P}(\r')$) are computed
numerically the asymptotic expansions of scattered fields become
analytic. This takes us to our
second task, which is 
making a direct connection between the orientation of
the induced polarization inside the defects and the far-field
radiation pattern, and in so doing define the scattering coefficients.\\
First of all, finding the scattered electric far-field $\E_s({\bf
R})$ requires the asymptotic expansions of the Green tensor. The
derivation is sketched in the Appendix~\ref{GT}. In what follows we
give some simplifying rearrangements that will let us focus directly
on the angular radiation pattern of surface defects.
\subsection{Scattering into Radiative Modes }
The asymptotic Green's tensor in the radiative zone  for either a
ridge or a groove can be written in a compact form as:
\begin{eqnarray} \label{asyp}\G^{(\nu)}(R\rightarrow\infty,\alpha,\r')
&=&  \,\frac{e^{i (g R + \pi/4 )}}{\sqrt{8\pi g
 R}}\,e^{-i g x'\cos\alpha}\nonumber\times\\[3mm]
  &\times&e^{-i k_z^\nu z'}\;\G_{\infty}^{(\nu)}(\alpha,\r').\end{eqnarray}
 In such form we can factor the asymptotic scalar green function out of the
dyadic part of the Green tensor.  From eq.(\ref{Es}), the direction
of $\E_s({\bf R})$ results from superposition of
$\G_{\infty}^{(\nu)}(\alpha,\r')\cdot{\bf P}^\nu(\r')$, the emission
from all induced point polarization elements, or dipole density
elements. Yet the direction of each contribution
$\G_{\infty}^{(\nu)}(\alpha,\r')\cdot{\bf P}^\nu(\r')\,$ must be
independent of $\r'$. In other words since electromagnetic waves are
transverse waves in vacuum, far from their source, the field emitted
by a dipole $\G_{\infty}^{(\nu)}(\alpha,\r')\cdot{\bf P}^\nu(\r')$
must be proportional to $\u_\alpha$.  In fact using the standard
asymptotic expansions (see the Appendix \ref{GT}) we can write:
\begin{eqnarray} \label{Inf}\G_{\infty}^{(\nu)}(\alpha,\r')\cdot{\bf P}^\nu(\r')=
 -\left[{\bf \Upsilo}^{(\nu)}(\alpha,\r')
\cdot{\bf P}^\nu(\r')\right]\;\u_\alpha .\end{eqnarray}
%
Where for a ridge:
\begin{eqnarray}{\bf \Upsilo}^{(1)}(\alpha,z') =
\k_p^+(\alpha)+\k_p^-(\alpha)\:r_p(\alpha)\:e^{2 i \g2 z' \sin\alpha
}.\end{eqnarray} and for a groove:
\begin{eqnarray}{\bf
\Upsilo}^{(2)}(\alpha)= t_p^{(1,2)}(\alpha)\, {\bf
k}_{pm}(\alpha).\end{eqnarray} The vectors $\k_{p}(\alpha)$ are
p-waves defined in vacuum, while $ \k_p^{m\pm}(\alpha)$ are defined
in the metal. A reminder of their expressions at normal incidence,
in terms of the angle $\alpha$ of Fig.~\ref{Scheme}, is reported in
the Appendix \ref{Pmodes}, along with the expression for the Fresnel
reflection
and transmission coefficients: $r_p(\alpha)$, $t_p^{(1,2)}(\alpha)$.\\
 We are now in a position to write the expressions
for the radiative fields. Plugging eq.(\ref{asyp}) and
eq.(\ref{Inf}) into eq.(\ref{Es}) we can  separate the electric far
field dependence into its radial and angular parts as:
\begin{eqnarray}{\E}_{s}^{(\nu)}(R,\alpha )=-\, \frac{e^{i(\g2 R - \pi/4)}
} {\sqrt{8 \pi \g2 R }}\;E_s^{(\nu)}(\alpha )\;\u_\alpha.
\end{eqnarray}
Here the angular amplitude can be written as:
\begin{eqnarray} E_s^{(\nu)}(\alpha )=g^2\,\sca^{(\nu)}(\alpha)\end{eqnarray}
where $\sca^{(\nu)}(\alpha)$ is the scattering coefficient into
radiative-modes:
\begin{eqnarray}
\sca^{(\nu)}(\alpha)=\int_{A}d\r'\, e^{-i g x'\cos\alpha}e^{-i
k_z^{(\nu)}z'} {\bf \Upsilo}^{(\nu)}(\alpha,\r') \cdot{\bf
P}^{(\nu)}(\r')\nonumber\\ \label{angularfield}
\end{eqnarray}
 In the last expression the scattered field in the far
zone consists of a cylindrical wave, transverse to the direction of
propagation $\u_R$, and with a net angular amplitude determined by
the integral over the source region $\sca^{(\nu)}(\alpha ) $. The
latter is actually the important bit in the formula as its squared
module determines the radiation pattern. As seen from
eq.(\ref{angularfield}) this angular amplitude results from the
superposition of each scattering element taken with its own
amplitude, phase and optical path in analogy to how an antenna array
determines its effective radiation pattern. The radiation is given
by the intensity or Poynting vector in the far field. Accordingly
the differential angular scattering cross-section is:
\begin{eqnarray}
 \label{sigmaout}\frac{\partial\sigma_{rad }^{(\nu)}(\alpha)}{\partial\alpha}&=&
\frac{|{\E}_{s}^{(\nu)}(R,\alpha)|^2
R}{S_{spp}}=\frac{g^3}{S_{spp}}|\sca^{(\nu)}(\alpha )|^2.\nonumber\\
\end{eqnarray}
Finally, the net radiative loss $\sigma_{rad}$ is defined as the
integrated angular radiation:
\begin{eqnarray}\label{net}\sigma_{rad}=\int_0^{180^0}d\alpha\;
\frac{\partial\sigma_{rad
}^{(\nu)}(\alpha)}{\partial\alpha}.\end{eqnarray}
\subsection{Shallow defects and Green's tensor boundary conditions }
Whenever the height of the defect is small enough, typically much
smaller than the wavelength of the incident light, we can make the
approximation $g|\r'|<<1$. That allows for some simplification
 for the angular amplitude 
of a scattering element \textit{{above the surface}}. Consider:
\begin{eqnarray}\label{Vector}{\bf \Upsilo}^{(1)}(\alpha,\r') &=&
(\k_p^+(\alpha)+\k_p^-(\alpha)\:r_p(\alpha)\:e^{2 i \g2 z'
\sin\alpha })\nonumber\simeq \\  \label{short} &\simeq&
(\k_p^+(\alpha)+\k_p^-(\alpha))\:r_p(\alpha)
\end{eqnarray}
Hence, for shallow defects the Green Tensor dependence of
eq.(\ref{asyp}) is entirely given by the exponential factors $e^{-i
g x'\cos\alpha}e^{-i k_z^\nu z'}$, for both a source in the vacuum
semi-space and a source in the metal semi-space. Indeed this turns
out to be a major simplification for the relative amplitude of the
scattering elements in the air
semi-space, which we shall perform in detail Section\ref{HV}.\\
Before that  we need to highlight the relation between the Green
tensor of a defect on the metal slab and in the metal slab, under
this approximation. Such relation emerges from the boundary
conditions for the Green's tensor at the interface, which are:
\begin{eqnarray}&{\label{Gsurfacex}}\left[\G({\bf R}, x',z'=0^+)-\G({\bf R},x',z'=0^-)\right]
\cdot\u_x=0&\nonumber
 \\\end{eqnarray}
\begin{eqnarray}
\label{Gsurfacez} &\left[\G({\bf R}, x',z'=0^+)-\varepsilon\,\G({\bf
R}, x',z=0^-)\right]\cdot\u_z=0.&\nonumber\\
\end{eqnarray}
Notice that, in the unperturbed system, space is translationally
invariant in the horizontal direction $x$ and this is reflected in
is the $x$-component of the vector in eq.(\ref{Vector}). Because of
eq.(\ref{asyp}) and eq.(\ref{Inf}), we can turn eq.(\ref{Gsurfacex})
into:
\begin{eqnarray}\Upsilo^{(1)}_{x}(\alpha)={\Upsilo}^{(2)}_{x}(\alpha)
={\Upsilo}_{x}(\alpha).\end{eqnarray}  The presence of surface
charges at the interface implies, from eq.(\ref{Gsurfacez}), that
the $z$-components of the vector ${\bf\Upsilo}^{(\nu)}(\alpha)$ on
either sides of the interface have the relation:
\begin{eqnarray}\label{Piz}\Upsilo^{(1)}_{z}(\alpha)=\varepsilon\;
{\Upsilo}^{(2)}_{z}(\alpha).\end{eqnarray}
\subsection{Scattering into Surface Plasmons}
Let us derive the scattering coefficient into surface plasmon modes.
Note that, in this one dimensional problem, scattering will be into
both the forward surface plasmon ${\bf e} _{spp+}(\r)$, propagating
in the positive $x$ direction and the backwards plasmon ${\bf e}
_{spp-}(\r)$ propagating in the negative
 $ x$ direction, as  defined in the Appendix \ref{SP}. The emission
 by a point dipole or a point polarization element
must result into a plasmon final state: $\G_{p\pm}({\bf
R},\r')\cdot{\bf P}(\r')\propto {\bf e }_{spp\pm}$, as shown in the
derivation sketched in the Appendix \ref{GT}.  The asymptotic Greens
tensor for a source \textit{upon} ($\nu=1$) or \textit{in} ($\nu=2$)
the metal is:
\begin{eqnarray}&
\G_{p\pm}^{(\nu)}({\bf R},\r')\cdot{\bf P}^{(\nu)}(\r')&=\frac{i}{2g
\,S_{spp}}
 \nonumber\times \\\label{Gp1}
 &\times\left[\,  \left({\bf e }^{(\nu)}_{spp\pm}(\r')\right)^*\cdot {\bf
P}^{(\nu)}(\r')\,\right]&{\bf e }_{spp\pm}({\bf R}).
\end{eqnarray}
Notice that $  \left({\bf e }^{(\nu)}_{spp\pm}(\r')\right)^*$
complies with eq.(\ref{Gsurfacex}) and eq.(\ref{Gsurfacez}).
Consequently the field of the scattered plasmons are:
\begin{eqnarray} \label{EP1} {\bf E}_{p}^{(\nu)\pm}&=& -\frac{i g}{2 \,S_{spp}}
\scap_{p\pm}^{(\nu)} \;{\bf e }_{spp\pm}\nonumber\\[5mm]
\scap_{p\pm}^{(\nu)}&=&
 \int_{A^{(\nu)}} d\r'\;{\bf e }^{*}_{spp\pm}(\r')\cdot {\bf P
}(\r')\end{eqnarray} Furthermore the magnetic field related to the
field scattered into SPPs is :
\begin{eqnarray}{\bf H}_{p\pm}=-\frac{i g}{2 \,S_{spp}}\, \scap_{p\pm}\;
{\bf h }_{spp\pm}\end{eqnarray} where ${\bf h }_{spp}$ is the
magnetic field of a SPP, as proved in the Appendix \ref{SP}.\\
Now, if the the source ${\bf P }(\r') $ is produced by an incident
surface plasmon field (as is our case), we can define the scattering
cross-section of into SPPs as:
\begin{eqnarray}\label{spmas}\sigma_p^\pm=\frac{{\bf E}_{p\pm}\times{\bf
H}_{p\pm}^*\cdot\u_x}{ {\bf e }_{p\pm}\times{\bf h}_{p\pm}^*
\cdot\u_x}= \left|\frac{ g}{2 \,S_{spp}}
\scap_{p\pm}\right|^2\end{eqnarray} Finally, we can define the total
scattering cross-section, which in the lossless case is equivalent
to the extinction cross-section:
\begin{eqnarray}\label{extn}\sigma_{xtn}=\sigma_p^++\sigma_p^-+\sigma_{rad}\end{eqnarray}
\section{\label{Ray}Rayleigh-limit: cautionary
remarks} Next we are going to develop solutions to point sources in
a metal plane background. However one question may be raised : how
do we associate the field induced by a surface plasmon inside a
ridge
or a groove to a point dipole? The answer is the argument of this section.\\
 When the field inside
a defect is obtained by mesh discretization we assume that the field
inside a single mesh is uniform, and deviations from the field at
its center are deemed negligible. Yet, in general, the field in a
defect, cannot be represented by the field at
its center alone. Let us explain a little bit further this point.\\
For simplicity let us consider a defect in a homogenous medium with
dielectric constant $\varepsilon_b $, but the argument is the same
in other backgrounds. As usual\cite{OJ}, the field at every mesh is
found by solving self-consistently a system of $N$ coupled equations:
\begin{eqnarray}\label{LS}{\bf E}(\r_i)= {\bf E}_b(\r_i)\;+\,g^2\,
\sum_{j\neq i}  \G_b(\r_i-\r_j)\cdot
\Delta\varepsilon \,{\bf E}(\r_j)\frac{A}{N^2}+\nonumber \\
\,+g^2\:{\bf \hat M}\cdot\Delta\varepsilon \,{\bf
E}(\r_i)-\frac{{\bf \hat L}}{\varepsilon_b}\cdot\Delta\varepsilon
\,{\bf E}(\r_i)\;\;\;
\end{eqnarray}where $i=1,N$ and $j=1,N$ and
${\bf E}(\r_i)$ is the field at the mesh center. ${{\bf \hat L}}$ is
a term related to the depolarization of light and comes about from
the quasi-static contribution of the Green tensor. ${{\bf \hat M}}$
is a correction term to the Green tensor in the region of the scatterer
useful to improve the accuracy of the calculation, when the inhomogeneity is discretized \cite{NanoOptics,MartinPiller}.\\
In practice, the number of mesh points $N$ is increased until the
calculation converges to the required precision. Then scale
variations $\sim \sqrt{A}/N$ of $\E(\r)$ are properly represented in
the solution. In the Rayleigh limit, for a defect of area $A$ so
small that $g^2 A<< 1$, the scatterer behaves like a point source or
a point dipole and the background field (in this case the
illumination) can be considered uniform over $A$: ${\bf
E}_b(\r)={\bf E}_b$ . Exceptionally, for a circular defect in a
homogenous medium with dielectric constant $\varepsilon_b $, the net
field at any point $\r_i$  converges to:
\begin{eqnarray}\label{Mie}{\bf E}= {\bf E}_b\; -\frac{{\bf \hat
L}}{\varepsilon_b}\cdot\Delta\varepsilon \,{\bf E}\;\;\;
\end{eqnarray}
\begin{figure}
\begin{center}
\includegraphics[scale=1,width=23pc,height=20pc]{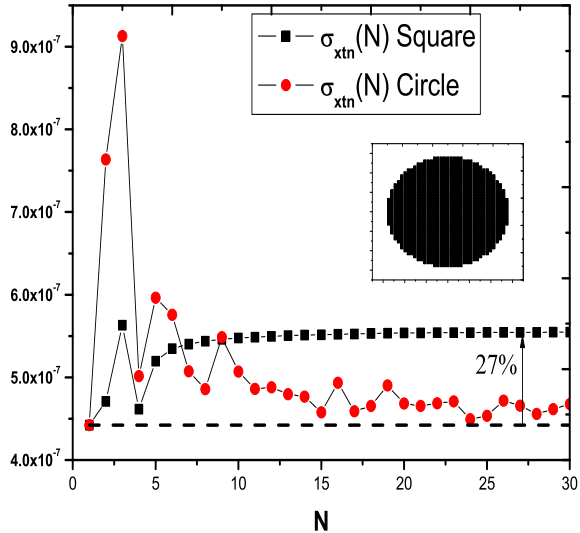}
\caption{Dependence of the extinction coefficient on the number of
meshes used in the calculation, for a square defect with $w=h=1nm$
in vacuum, and a circular defect of the same area, illuminated by a
plane wave. The dashed line represents extinction coefficient
calculated with the Mie theory for the circle. The dielectric
constant in the defect is $\varepsilon= - 19.89$ at the wavelength
of $700nm$. The inset represents the geometry of a discretized
circle when inscribed in a square represented by 30x30 mesh points.
}\label{Cylinder3}
\end{center}
\end{figure}
This is because for the field inside an infinitesimal (very
sub-wavelength) circular shape is actually uniform and thus
scattering by such circular defects can be described by one mesh. In
fact the  extinction coefficient\cite{Andrey,SonderBoz2003} can be
derived from the field at the center alone:
\begin{eqnarray}\label{xtn1}\sigma_{xtn}&=&g\Im\left[\int_A d\r'\,
\Delta\varepsilon \E^{*}_b(\r')\cdot\E(\r') \right]=\\
\label{xtn2}&=& A \Im\left[ \Delta\varepsilon \E^{*}_b\cdot\E
\right]
\end{eqnarray}
 To prove
this numerically  we have calculated  $\sigma_{xtn} $ for a cylinder
represented by a single mesh, as in eq.(\ref{Mie}), and illuminated
by a plane wave. First of all we have checked that the one-mesh
cross-section of eq.(\ref{xtn2}), coincides with the Mie theory
result. Secondly, we have subdiscretized the cylinder into
square meshes as rendered in the inset of Fig.~\ref{Cylinder3}. As
also rendered in the figure, applying eq.(\ref{LS}) we found that,
as the number of meshes grows, the scattering cross-section
calculated by the collection of meshes eq.(\ref{xtn1}) converges to
the initial value of one single mesh of eq.(\ref{xtn2}).
 However the field inside of a
square scatterer can never be uniform if it is to satisfy real
boundary conditions even in a homogenous medium or vacuum. Thus, it
can not be faithfully described by one mesh. This is illustrated in
Fig.~\ref{Cylinder3}, which renders the extinction coefficient for a
square defect of the same area as the circle.  As it turns out, the
converged value is $\sim 27\%$ larger than that obtained by the
one-mesh approximation. Remarkably this error is not reduced with
the defect size: we obtained the same error for squares with side
$5nm$ or $0.5nm$. This is just for reference in the optical range,
since we found that the error actually depends on type of defect and
on the dielectric constant.\\ However, even if the field is not
uniform, a small defect in the Rayleigh limit can be represented by
a point source at the center of the mesh, with its field equal to
the
average field over the mesh $\overline{\E}=(1/A)\int_A d\r'\,\E(\r') $.\\
Indeed if the variation of $\E_b(\r)$ is negligible over the area of the defect we
have:
\begin{eqnarray}\label{xtn3}\sigma_{xtn}=
A \Im\left[\Delta\varepsilon \E^{*}_b\cdot \overline{\E}\right]
\end{eqnarray}
So the object behaves as a point-dipole
$\p=A\Delta\varepsilon\overline{\E}$.\\ The previous results were
for a homogeneous background, but they also hold for the
inhomogeneous one considered in this paper. We find that, for a
defect above the surface in the optical range, the relative
error is about $40\%$, while it can reach $50\%$ for a defects below the surface.\\
  With very small non-elongated ridges and
grooves, such that $w/\lambda\approx h/\lambda<<1$, the equivalent point
dipoles are attained by averaging the fields over the area of the
defects as follows:
\begin{eqnarray}\label{SqrAverage}\p^{(1)}&=& \Delta\varepsilon \overline{\E}^r\, A=
\Delta\varepsilon \int_A d\r' \,\E^r(\r') \\\label{SqrAverage2}
   \p^{(2)}&=&-\Delta\varepsilon\overline{\E}^g \,A=-\Delta\varepsilon\int_A d\r'
    \,\E^g(\r')e^{-ig|z'|\sqrt{|\varepsilon|}}.\nonumber\\ \end{eqnarray}
Accordingly if we set ${\bf P}^{(\nu)}(\r)=\delta(\r-\r')\p^{(\nu)}$
 eq.(\ref{angularfield}) and eq.(\ref{EP1}) for small non-elongated defects become:
\begin{eqnarray}\label{p}\sca^{(\nu)}(\alpha)&=&{\bf
\Upsilo}^{(\nu)}(\alpha)\cdot\p^{(\nu)}\\[3mm]\scap_{p\pm}^{(\nu)}(\alpha)&=&
\left[{\bf e }_{p\pm}^{(\nu)}(0)\right]^*
 \cdot\p^{(\nu)}\end{eqnarray}
\section{\label{Num}Numerical results}
\begin{figure}
\begin{center}
\includegraphics[scale=1,width=20pc,height=13pc]{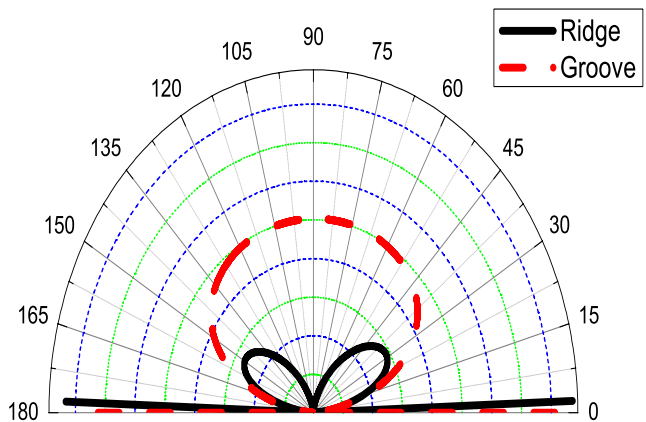}
\caption{\label{RadPatI}Angular radiative cross section
(eq.(\ref{sigmaout})) and surface plasmon cross section (eq.(\ref{spmas}) represented
by the almost horizontal lines at $\alpha\simeq$ and $\alpha\simeq
180^0$), for square defects with 10nm side, illuminated by a SPP on
silver at $ 500nm$. The scale is linear but the units are arbitrary.
 Each concentric line indicates an equal increment
of the cross-sections, from the minimum at the the center to the maximum at the outermost.
}
\end{center}
\end{figure}

As an illustration consider a square ridge and a groove of side
$w=h=10nm$. We have calculated the scattering into radiative modes
and SPPs without associating the defect to a point dipole but rather
using eq.(\ref{sigmaout}) and eq.(\ref{spmas}). In this case the
major task is computing the Green's tensor for the plane metal
surface required to attain the exact field
within the surface defect. This can be achieved following the prescriptions
of Ref.~[\onlinecite{Paulus,SonderBoz2004}].\\
 Similar numerical
results for the case of shallow grooves were found in
Ref.~[\onlinecite{Ioannis}] using a different computational
technique.
\\ The out of
plane radiation pattern of a surface plasmon scattered by such
defects is given in Fig.~\ref{RadPatI}.\\ Calculations show that,
for symmetric defects, the net radiative loss is greater for a
groove than for a ridge. This is so because, while both the
scattering into SPPs and the radiation close to the surface (at
$\alpha=0,180^\circ$) are similar, their radiation patterns greatly
differ normal to the surface ($\alpha=90^\circ$), where the groove
radiation is maximum while the ridge radiation goes to
zero.
\begin{figure}
\begin{center}
\includegraphics[scale=1,width=20pc,height=13pc]{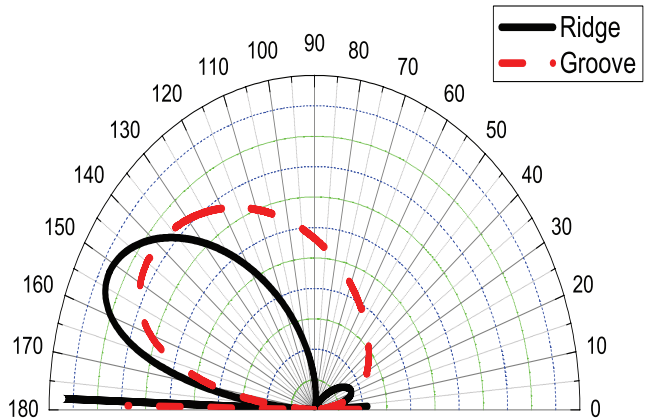}
\caption{ Angular radiative cross section (eq.(\ref{sigmaout})) and
surface plasmon cross section (eq.(\ref{spmas}) represented by the
almost horizontal lines at $\alpha\simeq$ and $\alpha\simeq 180^0$),
for rectangular defects with $10nm$ height and 50nm width,
illuminated by a SPP on silver at $ 700nm$. The scale is linear but
the units are arbitrary. Each concentric line indicates an equal
increment of the cross-sections, from the minimum at the the center
to the maximum at the outermost.
 } \label{Rect1}
\end{center}
\end{figure}
The ridge radiation pattern is distributed into two lobes on either
sides of $\alpha=90^\circ$ but the groove radiation pattern forms a
single lobe.  This is one of our main result and shall be analyzed in
detail in the next section. The result is not in agreement
with those obtained in the approximate treatment
Ref.[\onlinecite{Alexey}]. We associate the discrepancy to the
breakdown of the condition that the curvature of
a short and shallow defect does not vary rapidly, used in that work. \\
 Notice the fraction of energy scattered into SPPs,  i.e $\sigma_p^\pm $ of eq.(\ref{spmas}), is
large. The values of  $\sigma_p^\pm $ are represented by the horizontal lines of Fig.~\ref{RadPatI} (the concentric lines indicate their amplitude in a linear scale and in arbitrary units, say, for instance from $0$ at the center to $8$ at the outermost).
 For both ridges and grooves $\sigma_p^+ $ and $\sigma_p^- $ are roughly equal. However
in the case of ridges,  $\sigma_p^\pm  $ is greater than the maximum value of the scattering cross-section into radiative modes ${\partial\sigma_{rad }^{(1)}(\alpha)}/{\partial\alpha}$ of eq.(\ref{sigmaout}), by a factor slightly greater than $2$. For grooves, $\sigma_p^\pm $ is greater than the maximum value of ${\partial\sigma_{rad }^{(2)}(\alpha)}/{\partial\alpha}$ by a factor slightly smaller than $2$.\\
 Let us now keep the defects
height at $h=10nm$ and enlarge the width $w$. Fig.~\ref{Rect1}
renders the radiation pattern for a rectangular defect of width
$50$nm ($h=10nm$). The emergence of directivity in the out of plane
radiation, is part of a transitional behavior, in which the
radiation patterns tend to align and, simultaneously, one of the
lobes is shrunk while the other is blown up in the ridge radiation.
Notice that the scattered energy into SPPs exhibits the same
directivity, going mainly in reflection. Eventually, if we keep
enlarging the defects until they are considerably asymmetric the
radiation patterns for both ridges and grooves tend to be single
overlapping lobes (see Fig.~\ref{Rect2}). Noticeably, the scattering
into SPPs is greatly reduced. Such similarity is explainable in the
approximate framework presented in Ref.[{\onlinecite{Alexey}}] which
turns out to be quite acceptable in this limit of large enough
defects, as we shall substantiate in Sec.~\ref{NeedleSection}.
In Sec.~\ref{Transition} we shall account qualitatively for the
transition observed in Fig.~\ref{Rect1}, explaining why the
radiation pattern changes when the defects are enlarged.
\begin{figure}
\begin{center}
\includegraphics[width=20pc,height=13pc]{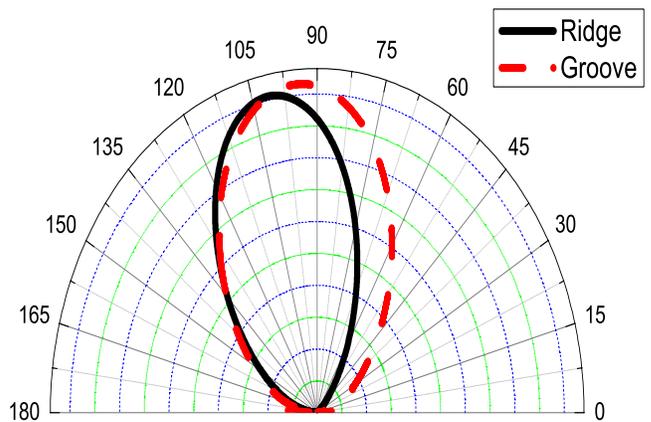}
\caption{ Angular radiative cross section (eq.(\ref{sigmaout})) and
surface plasmon cross section (eq.(\ref{spmas}) represented by the
almost horizontal lines at $\alpha\simeq$ and $\alpha\simeq 180^0$),
for rectangular defects with $10nm$ height and $300nm$ width,
illuminated by a SPP on silver at $700nm$. The scale is linear but
the units are arbitrary. Each concentric line indicates an equal
increment of the cross-sections, from the minimum at the the center
to the maximum at the outermost. \label{Rect2}}
\end{center}
\end{figure}
\subsection{Scattering by square ridges and grooves in the Rayleigh limit}
\begin{figure}
\begin{center}
\includegraphics[scale=1,width=20pc,height=15pc]{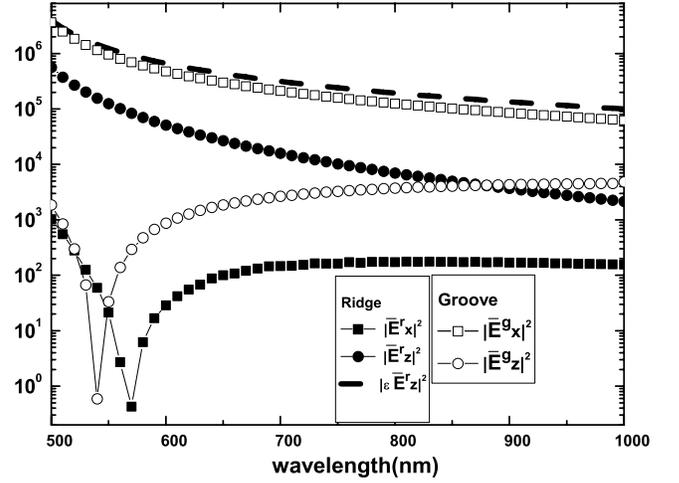}
\caption{The averaged field components (as defined in
eq.(\ref{SqrAverage}), eq.(\ref{SqrAverage2})) for a square groove
and a square ridge of 10nm side in silver, as a function of the
wavelength. The scale is logarithmic with arbitrary
units. }\label{ModE}
\end{center}
\end{figure}

The equivalence between non-elongated subwavelength defects and
point dipoles gives us a chance to investigate in depth the
individual radiation pattern of a single scattering element.\\
Fig.~\ref{ModE} shows the averaged the field inside the 10nm ridges
and grooves, as prescribed in eq.(\ref{SqrAverage}) and
eq.(\ref{SqrAverage2}).   The field induced in a groove is mainly
longitudinal while the field inside the ridge is mainly transversal.
This is due to both the illumination and the polarizabilty of the
scatterers. When defects are almost symmetric their polarizibilities
$\beta_i$ are nearly isotropic and so the induced field and the
incident field are virtually parallel. Hence the field induced in a
ridge and a groove are nearly parallel to the incident surface
plasmon ${\bf e}_{spp}$, which is mainly perpendicular to the plane
in the vacuum semi-space and is mainly parallel to the plane in the
metal semi-space. Therefore, in the Rayleigh limit, a ridge scatters
SPPs into radiative modes like a vertical dipole on the plane, while
the groove scatters them into radiative modes like a horizontal
dipole on a plane. The results for grooves is in agreement with
Ref.[\onlinecite{LevDipole}].
\\Interestingly, we also have
found numerically in Fig.~\ref{ModE} that:
\begin{eqnarray}\label{Exz} |\overline{E}_x^g|\sim|\sqrt{\varepsilon}\;\; \overline{E}_{z}^{r}|
\end{eqnarray} especially at short wavelengths. We have devised a virtual source,
 that can condense the orientation of the equivalent dipole representing
 a non-elongated symmetric
ridges and grooves. This virtual dipole is defined as: ${\bf
q}(\theta)=\u_x\sqrt{|\varepsilon|}\cos\theta+\u_z \sin\theta$. The
fields inside a groove and a ridge, are respectively, represented 
as:\begin{eqnarray} \label{apprsim}|\overline{\E}^{g}|&\simeq&
|\Delta\varepsilon\,\overline{E}^r_{z}{\bf q}(0)|
\\[4mm] | \overline{\E}^{r}|&\simeq& |\Delta\varepsilon\,\overline{E}^r_{z}{\bf q}(90^0)|
\end{eqnarray}
at least as long as eq.(\ref{Exz}) holds.\\
In reality we can see what happens by means of eq.(\ref{Mie}).
Despite the fact that this equation is only exact for a circle in a
homogenous background (as explained) we can use it to show
 \textit{qualitatively} the relation between the field inside the groove and
the ridge, when their shapes are symmetric. If we approximate the
polarizability of a ridge for that of a circle in vacuum (whose
polarizability is calculated through eq.(\ref{Mie})), so
$\beta_1=2/(\varepsilon+1)$. If we also approximate the groove
polarizability by that of a hole in a homogenous metal medium, we
have: $\beta_2=2\varepsilon/(\varepsilon+1)$. Hence the field
induced inside each object is:
\begin{eqnarray}\overline{\E}^r&\approx&\beta_1{\bf
e}_{spp}(x=0,z=0^+)= \beta_1\u_z \\
 \overline{\E}^g&\approx&\beta_2
{\bf e}_{spp}(x=0,z=0^-)= \beta_2\frac{\u_x}{\sqrt{\varepsilon}}
\end{eqnarray}  Since these polarizabilities also have the
property: $\beta_2=\varepsilon\beta_1$ (the polarizability of a hole
in a material is $\varepsilon$ times larger than the polarizability
of a particle of the same material and the same shape) then
$|\sqrt{\varepsilon}\; \overline{E}^g_{x}|\sim |\overline{E}^r_{z}|
$.
\\ The symmetry of the polarizations $\beta_i$ and the property
$\beta_2\simeq\varepsilon\beta_1$ are strictly true for circular
defects in homogeneous media. Our numerical calculations of
Fig.~\ref{ModE} shows that, even though the field inside a ridge and
a groove are quantitatively different from those of circular defects
in homogenous media, the assumption that their mutual relation is
preserved is in very good agreement with the exact result. Because
of the symmetry of the square shape, the \textit{averaged field}
inside the square is very nearly parallel to the incident field.
\subsection{Reflection of surface plasmons square shallow defects}
 As a corollary of the properties of the fields in a
ridge and a groove $|\sqrt{\varepsilon} \overline{E}^r_{z}|\sim
|\overline{E}^g_{x}| $ we can also substantiate that their
reflection of surface plasmons is quite similar. In fact, we obtain:
\begin{eqnarray}|E_{p\pm}^{(1)}|\simeq |\overline{E}^r_{z}|  \end{eqnarray}
\begin{eqnarray}|E_{p\pm}^{(2)}|\simeq
 \left|\frac{\overline{E}^g_{x}}{\sqrt{\varepsilon}}\right|\simeq
  |{\overline{E}^r_{z}}|\simeq |E_{p\pm}^{(1)}| \end{eqnarray}
Notice that these define $\sigma_p^{\pm}$ through eq.(\ref{spmas}).
Once $\sigma_{rad}$ from eq.(\ref{net}) and $\sigma_p^{-}$ are
determined the value of the transmission of the surface plasmon is a
constrained variable: $T=1-\sigma_p^{-}- \sigma_{rad}$, at least for
the lossless case{\cite{Alexey}}. Since $\sigma_{rad}$ is greater
for grooves than for ridges, the groove transmission is smaller.
 \section{\label{HV} Radiation
patterns for Horizontal and Vertical point dipoles on a real metal
interface} The first part of the expression eq.(\ref{sigmaout}) is a
pre-factor ${g^3}/{S_{spp}}$ whereas the second part is the
\textit{the radiation pattern} of a point dipole:
\begin{eqnarray}\label{rp}|\scap^{(\nu)}|^2=\left| {\bf\Upsilo}^{(\nu)}(\alpha)\cdot\p
\right|^2\end{eqnarray} A groove emits like a horizontal dipole. The
angular amplitude of the field radiated by a horizontal unit dipole
 $\p=\u_x$, placed close to the interface $z=0$, is ${\Upsilo}_{x}(\alpha)$,
and it does not matter on which side of the interface it is placed.
${\Upsilo}_{x}(\alpha)$ can be derived using the relations in the
Appendix \ref{Key} and the explicit result is:
\begin{eqnarray}\label{Gammax}
 {\Upsilo}_{x}(\alpha)=\frac{2
\sqrt{\varepsilon-\cos^2\alpha}\;\sin \alpha}
            {\sqrt{\varepsilon-\cos^2\alpha}+\varepsilon \sin\alpha}\end{eqnarray}
and the radiation pattern is $|{\Upsilo}_{x}(\alpha)|^2$. Notice
${\Upsilo}_{x}(\alpha)$ presents a mirror symmetry about the angle
$\alpha=90^0$, the normal to to the plane. Furthermore since
${\Upsilo}_{x}(\alpha)$ never changes sign between $0$ and 180$^0$
(nor goes to zero), the field of a horizontal dipole has one single
symmetric lobe, where the field always has the same
sign.\begin{figure}
\begin{center}
\includegraphics[scale=1,width=20pc,height=13pc]{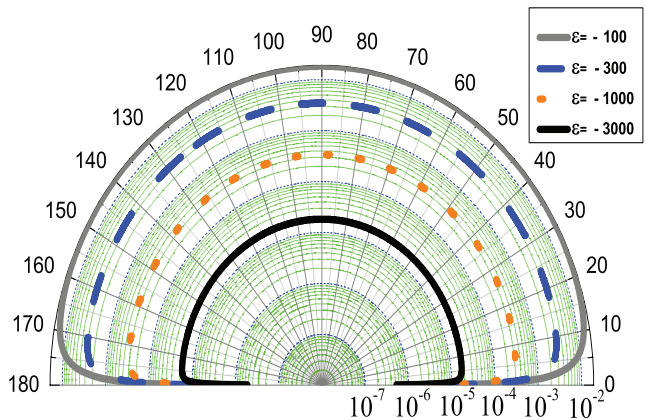}
\caption{Radiative angular intensity $|\Upsilo_x |^2$ of a
horizontal point dipole at an air-metal interface. The radiation
patterns varies as the metal dielectric constant is varied. The
scale is logarithmic with arbitrary units. }\label{ux}
\end{center}
\end{figure} The field intensity
$|{\Upsilo}_{x}(\alpha)|^2$ of such lobe is rendered in
Fig.~\ref{ux} for different dielectric constants. This radiation
pattern of a groove shown in Fig.~\ref{ux}, is in agreement with the
one represented by Ref.[\onlinecite{Ioannis}], obtained with a
different numerical method. Notice that for $|\varepsilon|>>1$:
\begin{eqnarray} {\Upsilo}_{x}
\rightarrow 2\;{{\varepsilon}^{-1/2}}.\end{eqnarray}That is, when
$\varepsilon$ increases this radiation pattern tends to become
simultaneously isotropic and vanishing. In fact a horizontal dipole
does not radiate on a perfect conductor\cite{SommerfeldBook}. On a
small digression it is interesting to notice an apparent
contradiction between treatments such as Ref.[\onlinecite{Bethe}],
which considered that a defect in a perfect metal were equivalent to
a magnetic dipole, while another work\cite{LevDipole} explains a
defect in a real metal corresponds to an electric dipole. Actually
we have just reconciled the two results. We know that a horizontal
dipole on a plane tends to emit isotropically for large
$\varepsilon$. This means that on a first order expansion in
$1/\varepsilon$, the radiation pattern of a horizontal dipole on a
plane and that of a magnetic dipole in vacuum, are identical.\\ For
finite $\varepsilon$ the field ${\Upsilo}_{x}(\alpha)$ of a
horizontal dipole within a real metal would not be thoroughly
screened, and while the pattern remains symmetric, its isotropy is
disrupted parallel to the surface (i.e. $\alpha=0,180^0$) to
accommodate the emergence of the surface
plasmons density of states. \\
For an individual vertical dipole $\p=\u_z$, which represents a ridge,
 the angular amplitude of
the field is (see Appendix \ref{Key}) :
\begin{eqnarray}\label{Gammaz}
 {\Upsilo}_{z}^{(1)}(\alpha)=\frac{2 |\varepsilon| \sin \alpha}
            {\sqrt{\varepsilon-\cos^2\alpha}+\varepsilon \sin\alpha}\;\;\large\cos\alpha\end{eqnarray}
 The field from a vertical dipole also goes to zero at
$\alpha=0,180^0$  for a finite $\varepsilon $, but since dipoles
only radiate transversally, the field has a third zero at $90^0$.
The field is antisymmetric with respect to the normal of the plane,
while the intensity $|{\Upsilo}_{z}(\alpha)|^2 $ is symmetric, and
is made up of the two lobes separated by a zero at 90$^0$, see
Fig.~\ref{symmetric}. Yet it is important to keep in mind that the
field of one lobe is in anti-phase with the field of the
other.\begin{figure}
\includegraphics[scale=1,width=20pc,height=13pc]{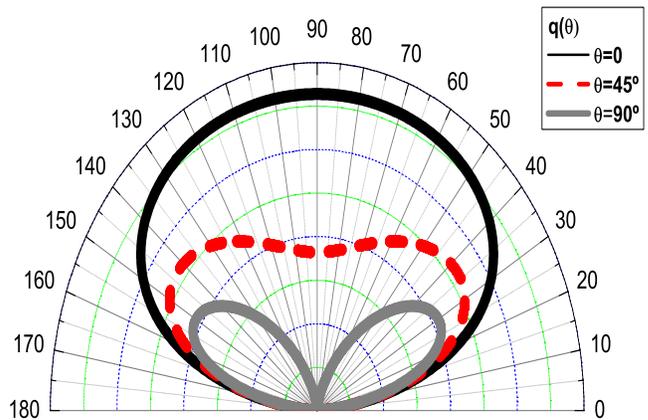}
\caption{Radiation pattern $|\sca(\alpha)|$ from the virtual dipole
$q(\theta)$ at $\theta=0,45^0,90^0$. The scale is linear with
arbitrary units. }\label{symmetric}
\end{figure}
\\Unlike a horizontal dipole, the vertical dipole radiative field
does not vanish for $|\varepsilon|>>1$ in fact:
\begin{eqnarray}
{\Upsilo}_{z}^{(1)} \rightarrow 2 \cos\alpha
\end{eqnarray}
The total radiation from a vertical dipole has a larger weight than
the radiation by a horizontal one, by a factor of
$\sqrt{\varepsilon} $. This can be seen, in fact, from
eq.(\ref{Gammax}) if  we assume $\varepsilon $ is large, we get the
following relation:
\begin{eqnarray}\label{uguale}\Upsilo_{z}^{(1)}(\alpha)
\simeq\sqrt{\varepsilon}\,\;{\Upsilo}_{x}(\alpha)\cos\alpha
\end{eqnarray}
 In Fig.~\ref{symmetric} we represent radiation pattern of ${\bf q}(\theta)$ for the horizontal
 and vertical orientations respectively, $\theta=0$,
$\theta=90^0$, which corresponds to our analytic analog of the
emission pattern of square ridges and grooves respectively. While we
will consider an intermediate orientation in the next section, we
want to remark here that, due to eq.(\ref{uguale}), the radiation by
both the horizontal moment ${\bf q}(0)$ and a vertical moment ${\bf
q}(90^0)$  vanish parallel to the plane at $\alpha=0,180^0$ in a
similar manner,
as illustrated in Fig.~\ref{RadPatI} \\
At the same time the far-field emissions of ridges and grooves
become increasingly
different as we approach the direction normal to the plane. 
\section{\label{NeedleSection}Solutions for long and shallow Ridges and Grooves }
For  shallow and long defects $w>h$ and $ h/\lambda<<1$ we define
the following height-averaged polarization densities  and fields:
\begin{eqnarray} {\bf \widetilde{P}}^{(1)}(x')&=&\Delta\varepsilon \int_0^h
dz'\E^{(1)}(x',z')= \nonumber\\ &=&
\Delta\varepsilon\;\widetilde{\E}^{(1)}(x')\; h\end{eqnarray} where
the last equation defines ${\widetilde{\E}}^{(1)}(x')$. Likewise for
a groove we can define ${\bf \widetilde{P}}^{(2)}(x')$ and
$\widetilde{\E}^{(2)}(x')$ through the following equation :
\begin{eqnarray} {\bf \widetilde{P}}^{(2)}(x')&=&-\Delta\varepsilon\int_{-h}^0
dz'\E^{(2)}(x',z')\;e^{-g|z'|\sqrt{|\varepsilon|}}\nonumber\\
&=&-\Delta\varepsilon\; \widetilde{\E}^{(2)}(x')\;h\end{eqnarray}
Notice for $|\varepsilon|>>1$  we can make the approximation
$k_{pz}^m\sim k_z^m\sim i g\sqrt{\varepsilon} $.\\
The benefit of using $ {\bf \widetilde{P}}^{(\nu)}(x')$ is that the
scattered-field coefficients for these defects in the far zone,
$\sca^{(\nu)}(\alpha)$ and $ \scap_{p\pm}^{(\nu)}$, are those emitted
by a chain of point-dipoles on the surface over
the segment $w$, and set  at $0^+$ and $0^-$ for ridges and grooves, respectively.\\
 The scattered field angular amplitude $\sca^{(\nu)}(\alpha
)$ from eq.(\ref{angularfield}) and  eq.(\ref{short}) is obtained
as:
\begin{eqnarray}\sca^{(\nu)}(\alpha)
&\simeq& {\bf \Upsilo}^{(\nu)}(\alpha)\cdot \int_{0}^w d\r'\,{\bf
\widetilde{P}}^{(\nu)}(x')\, e^{-i g x'\cos\alpha}\end{eqnarray}
\begin{figure}
\begin{center}
\includegraphics[scale=1,width=20pc,height=15pc]{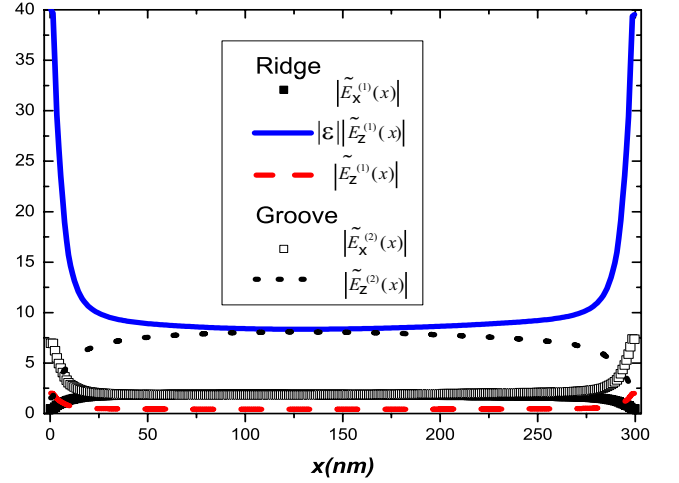}
\caption{\label{FieldLong} The averaged fields component inside of a
ridge and a groove, $\widetilde{E}^{(1)}_x$, the quantity
$\varepsilon\widetilde{E}^{(1)}_z$ and the rest of the components
$\widetilde{E}^{(1)}_x$, $\widetilde{E}^{(2)}_x$,
$\widetilde{E}^{(2)}_z$, for rectangular defects of $w=300nm$ and
$h= 10nm$. The system is illuminated by a SPP in lossless silver at
$\lambda=700$nm. The scale is linear with arbitrary units.}
\end{center}
\end{figure}
This holds for the scattering into surface plasmon modes as well
since we have:
\begin{eqnarray}
 \scap_{p\pm}^{(\nu)}&=&
  \left[{\bf e}_{spp\pm}^{(\nu)}(0)\right]^*\cdot
 \int_0^w dx'\;e^{\mp i k_{px}x'}\;
{\bf \widetilde{P}}^{(\nu)}(x').\end{eqnarray}
When we illuminate a shallow and long defect, with a SPP, an
equivalent linear density of dipole sources ${\bf\widetilde{
P}}(x')$  stems from how the induced fields are distorted inside the
scatterer, namely by its polarizability. When the defect is larger
in the horizontal direction than in the vertical one, ridges and
grooves were found to give the same scattering by an approximated
Rayleigh expansion\cite{Alexey}. We have an alternative first
principles argument to justify the Rayleigh expansion result, which
is based entirely on the assumption that these defects are needle
shaped. The field induced in these defects tends to be that induced
in a needle-shaped protrusion placed horizontally on the surface
$0^+$ in the case of a ridge. For a groove we have a horizontal
needle-shaped cavity at $0^-$. In such idealistic simplification it
is clear-cut to deduce the fields inside the defects from the
boundary conditions. Namely the parallel component of the  incident
field is always continuous and equal, as in eq.(\ref{Ispp1})and
eq.(\ref{Ispp2}):
\begin{eqnarray}\widetilde{E}_{1x}(x')={\bf
e}_{sp}(x',0)\cdot\u_x=\widetilde{E}_{2x}(x')\end{eqnarray} which
preserves the continuity of eq.(\ref{Gsurfacex}). However, we are
generating fields which, normal to the surface, make up for the
discontinuity perpendicular to the metal surface of
eq.(\ref{Gsurfacez}). In fact, for a horizontal needle-like ridge,
the boundary conditions imposed by the continuity of the
displacement vector are:
\begin{eqnarray}\widetilde{E}_{1z}(x')={\bf
e}_{spp}(x',0^+)\cdot\u_z/\varepsilon=1/\varepsilon\end{eqnarray}
while for a needle-like slit:
\begin{eqnarray}\widetilde{E}_{2z}(x')=\varepsilon\;{\bf
e}_{spp}(x',0^-)\cdot\u_z=1.\end{eqnarray} Ultimately:
\begin{eqnarray}\label{needle}
\widetilde{E}_{1x}(x')&=&\widetilde{E}_{2x}(x')\\\label{needle2}
\varepsilon\widetilde{E}_{1z}(x')&=&\widetilde{E}_{2z}(x')
\end{eqnarray} which,  matched with eq.(\ref{Gsurfacex}) and eq.(\ref{Gsurfacez}), yields:
\begin{eqnarray}\label{alex}&\left|\G({\bf R}, x',z'=0^+)\cdot \widetilde{\E}_1(x')\right|=&\\
&=\left|\G({\bf R}, x',z'=0^-)\cdot
\widetilde{\E}_2(x')\right|&\nonumber
\end{eqnarray}
and thus the property of producing the same scattering coefficients,
previously found in Ref.[\onlinecite{Alexey}]. Of course this is
just an approximation, but it explains why elongated defects have
similar scattering properties. In real life the plasmon scattering
by protrusions and indentations is similar because, far from the
edges, a shallow but elongated defect behaves as an infinitely
elongated one, as confirmed by numerical calculations. As an example
we report in Fig.~\ref{FieldLong} a numerical calculation of the
fields averaged over the height  for defects of $w=300nm$ and
$h=10nm$. This shows that eq.(\ref{needle}) and eq.(\ref{needle2})
are quite accurate at the center of the defect, and deviate from
the needle model prediction due to fringe effects at the edges.\\ It
is worth mentioning that this equivalence is valid in the Rayleigh
limit when the defect size is much smaller than the wavelength, and
may be altered at resonant wavelengths.
 \section{\label{Transition}The transition from short and shallow
defects to long and shallow defects: oblique dipoles on a real metal
plane} Everything we just said for symmetric surface defects was
based on the fact that their aspect ratio equals one. As the defect
width is increased,
 the aspect ratio becomes larger and this leads, progressively, to an
 asymmetric
polarizability tensor. The first effect is that the field induced is
gradually less and less parallel to the incident field. Therefore a
ridge would develop a non-negligible horizontal electric field
component, thus ceasing to be equivalent to a vertical dipole.
Likewise the groove, which in the symmetric case behaves as a
horizontal dipole, gradually starts having a non-negligible vertical
component  as its shape is elongated. The process goes on until we
recover the case of a needle shaped defect of section
\ref{NeedleSection}. The fields inside a defect having intermediate
width, as in Fig.~\ref{Rect1}, are intermediate between those for
the needle case and the square symmetric case. Therefore in these
cases defects emit qualitatively like \textit{oblique dipoles}, with
the orthogonal components out of phase.\\In order to understand
better the radiation pattern by ridges and grooves we decompose the
oblique dipole in its horizontal and vertical components.\\ First of
all, we focus on the mechanisms involved radiation pattern for a
ridge $\nu=1$. From eq.(\ref{p}) a dipole with arbitrary orientation
emits close to the surface, with a field angular amplitude:
\begin{eqnarray}\label{ridgeRad}\sca^{(1)}(\alpha)={\bf \Upsilo}^{(1)}(\alpha)
\cdot{\bf p}^{(1)}&=&
{\Upsilo}_{x}(\alpha)\Delta^{(1)}(\alpha)\end{eqnarray} where
$\Delta^{(1)}(\alpha)=
p_x^{(1)}+\left({\Upsilo}_{z}^{(1)}(\alpha)/\Upsilo_{x}(\alpha)
\right)\,p_{z}^{(1)}$ and equals:
\begin{eqnarray}
\Delta^{(1)}(\alpha)= p_{x}^{(1)}+
 i\, \frac{ \varepsilon
\cos\alpha}{\sqrt{\cos^2\alpha+|\varepsilon|}}\,p_{z}^{(1)}
\end{eqnarray}
$\Delta^{(1)}(\alpha)$ shows that the contribution to the radiative
field coming from the vertical and horizontal dipole on a metal
plane have a phase difference of $90^0$. This was already evident
from eq.(\ref{uguale}), when $\varepsilon<0$. Such phase difference
arises from the impedance of a metal plane\cite{Alexey}
$Z_s=-i/{\sqrt{|\varepsilon|}}$.\\ The radiation pattern for a
dipole with arbitrary orientation and lying above the metal,
is written in our formalism as:
$|{\Upsilo}_{x}(\alpha)\Delta^{(1)}(\alpha)|^2$.
\begin{figure}
\begin{center}
\includegraphics[scale=1,width=20pc,height=13pc]{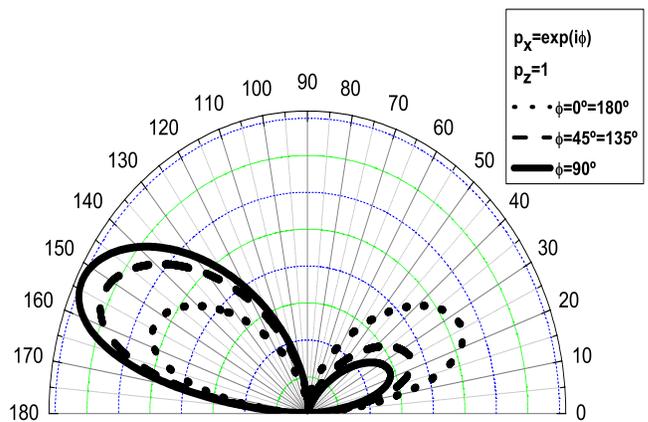}
\caption{Radiation pattern $|\sca(\alpha)|$ for a point dipole:
$\p=\u_x e^{i\phi}+\u_z$, lying on top of a metal surface. The scale
is linear with arbitrary units. }\label{dipoles2}
\end{center}
\end{figure}  The net angular amplitude for an oblique dipole is resolved into the
superposition of the angular envelope of the horizontal dipole
(shown in Fig.~\ref{ux}), with the other radiation factor
$|\Delta^{(1)}(\alpha)|^2$. This last factor contains both the
orientation and phase of the field. To envisage how these combine we
may develop
 $|\Delta^{(1)}(\alpha)|^2$ into three terms. These consist in
 the individual emission from the horizontal and vertical
dipole plus an interference term:
\begin{eqnarray}\label{plane}  |\Delta^{(1)}(\alpha)|^2 = |p_{x}|^2
+\frac{|\varepsilon|^2\,\cos^2\alpha}{|\varepsilon|+\cos^2\alpha}|p_{z}|^2+
\nonumber
\\
-2
\;\frac{|\varepsilon|}{\sqrt{\cos^2\alpha+|\varepsilon|}}\;\Im\left[p_{1x}
p_{z}^* \right]\cos\alpha &  \end{eqnarray} In the presence of the
plane metal background, we have that horizontal and vertical dipoles
behave as individual sources but their interaction presents an
intrinsic added phase difference of $90^0$, which is due to the
different interaction of a horizontal and a vertical dipole with the
plane.
 As a result, when in phase they do not interfere, and their
radiation pattern is always symmetric regardless of the orientation
of the dipole. This is the case for ${\bf q}(45^0)$ where, as in
Fig.~\ref{symmetric}, the radiation pattern is the sum of the
angular intensity of a vertical and a horizontal dipole, so that at
$90^0$ there is a minimum due to the vanishing of the vertical
dipole contribution, and yet never goes to zero because of the
horizontal dipole contribution. Nevertheless, when the dipole
components are not in phase, we can get asymmetric radiation
patterns and additional zeros (to those at $0^0$ and $180^0$),
because the interaction term can be negative. In such case the
interaction of the horizontal radiative field (with only one lobe)
with the vertical radiative (with two lobes of different sign) is
responsible for an asymmetric radiation pattern and exhibits
directionality.
\begin{figure}
\begin{center}
\includegraphics[scale=1,width=20pc,height=13pc]{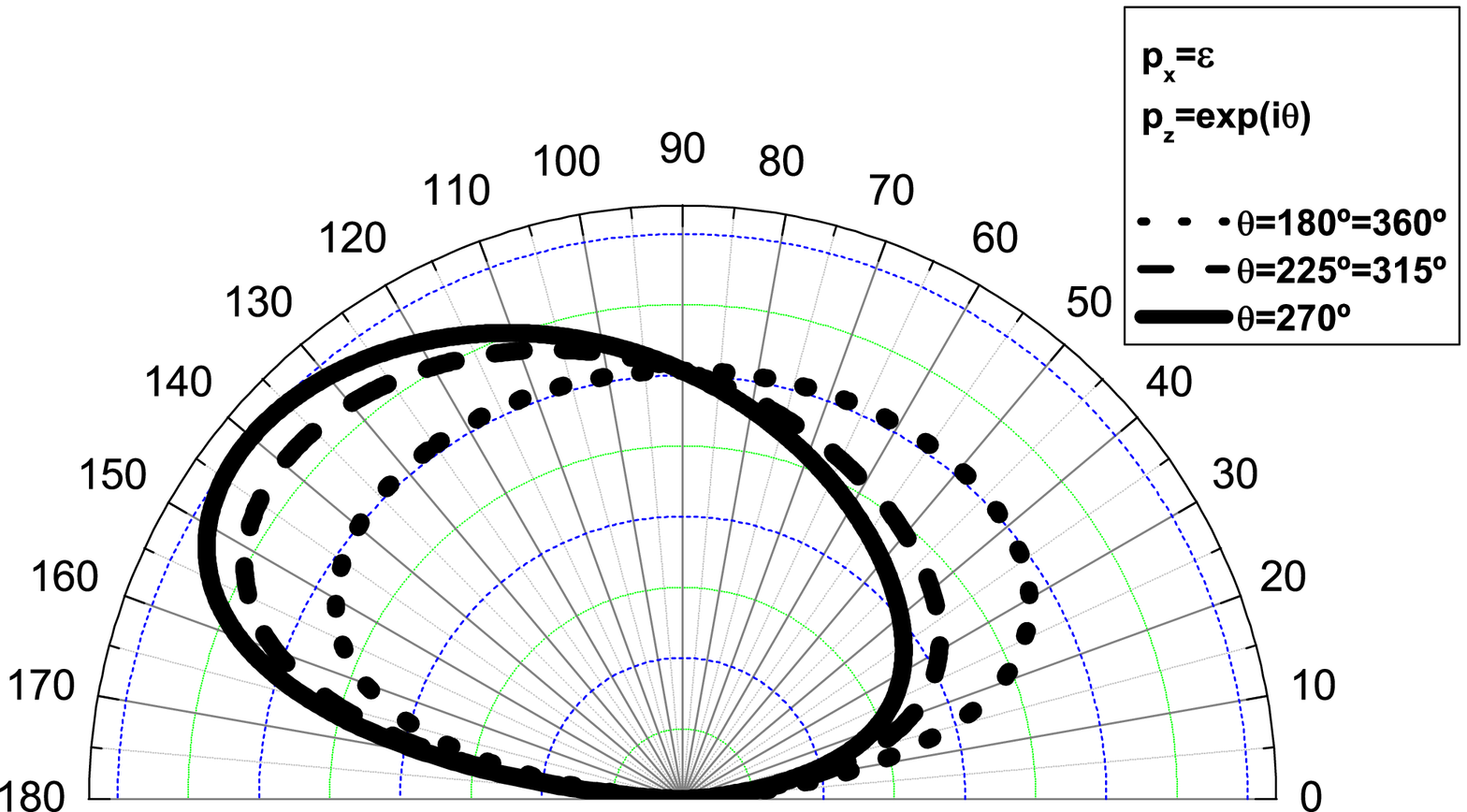}
\caption{Radiation pattern $|\sca(\alpha)|$ for a point dipole:
$\p=\varepsilon\u_x+ e^{i\phi}\u_z$, lying on top of a metal
surface. The scale is linear with arbitrary units.}\label{dipoles3}
\end{center}
\end{figure}
This is illustrated in Fig.~\ref{dipoles2} for a dipole emission
whose main contribution comes from the vertical dipole. In
Fig.~\ref{dipoles3} we show the radiation pattern for a dipole whose
main contribution comes from the horizontal dipole radiation.\\ For
the case of a grooves ($\nu=2$), the radiative angular field
amplitude is, from eq.(\ref{Piz}):
\begin{eqnarray}\sca^{(2)}(\alpha) &=&\Upsilo_x(\alpha)
\Delta^{(2)}(\alpha)\;e^{-g|z'|\sqrt{|\varepsilon|}}\\[3mm]
\Delta^{(2)}(\alpha)&=&p_{x}^{(2)}+ i\; \frac{
\cos\alpha}{\sqrt{\cos^2\alpha+|\varepsilon|}}\;p_{z}^{(2)}\end{eqnarray}
where remember we have also added the approximation: $ k_z^m \simeq
g\sqrt{\varepsilon}$ for $|\varepsilon|>>1$.\\ Remarkably, as
opposed to the the dipole emission over the surface, in the net
emission from a dipole under the surface the horizontal dipole
contribution has a greater weight than the vertical dipole
contribution. Apart from this, all the arguments used for a dipole
over
the surface apply.\\
The interaction between the vertical and horizontal components of
the field induced in the field generates the directional patterns of
Fig.~\ref{Rect1}. For a ridge with length slightly larger than its
height the directional radiation is dominated by its vertical
component. Fig.~\ref{dipoles2} exemplifies the effect of the
interference of a dominant vertical component with a smaller but
non-negligible horizontal component. For even larger aspect ratios
the contribution from the other component may be comparable.\\
Likewise when a groove has a small aspect ratio it is predominantly
a horizontal source interfering with a smaller vertical source. The
result is in an interference pattern that looks like the one
rendered in Fig.~\ref{dipoles3}. Yet again this can be modified by
increasing the aspect ratio. This transition is in good agreement
with Fig.~11 of Ref.[\onlinecite{Ioannis}] where, using a different
numerical method, the radiation pattern of a groove was computed for
different aspect ratios.

\section{Conclusions}
Our analysis of the surface plasmon scattering by square shallow
defects into radiative modes and plasmon modes, reveals that a
groove scatters more of the incident energy than a ridge does. The
reflection by a symmetric ridge and a groove is similar and so is
the radiative emission close to the horizontal direction. Indeed
their scattering essentially differs in the vertical direction,
where a groove scatterers while a ridge does not. 
  When defects start to
become longer in width we saw the polarizability gets more
asymmetric. Correspondingly, since both components of the incident
plasmon are out of phase, defects are equivalent to interfering
horizontal and vertical dipoles on a plane, which interfere
constructively in some direction, thus producing directionality in
the radiation pattern. Finally when ridges and grooves are shallow
and long they tend to produce the same scattering as, apart for
fringe effects, their polarizability exactly counterbalances the
discontinuity of the incident surface plasmon field at the air-metal
interface.

\section{Acknowledgments}
The authors acknowledge financial support from the Spanish Ministry
of Science and Innovation under grants NO.AP2005-5185,
MAT2008-06609-C02 and CSD2007-046-Nanolight.es.
\appendix
\section{Surface Plasmon Polariton Mode \label{SP}}
The incident illumination is the field of a surface plasmon wave
mode propagating in the positive x direction $(+)$ or negative x
direction $(-)$ is:
\begin{eqnarray}\label{Ispp1}{\bf e }_{spp\pm}^{(\nu=1)}(\mathbf{r})=\left(
\frac{\pm\u_x}{\sqrt{\varepsilon}}+ \u_z \right)  \;   e^{i (\pm
k_{p}x+k_{pz}z)}, \;\; \;z>0,\;\;
\end{eqnarray}
\begin{eqnarray}\label{Ispp2}{\bf e }_{spp\pm}^{(\nu=2)}(\mathbf{r})
=\left(\frac{\pm\u_x}{\sqrt{\varepsilon}}+ \frac{\u_z}{\varepsilon}
\right)  \;   e^{i
 (\pm k_{p}x+k_{pz}^m|z|)},\;\;\;z<0.\;\; \end{eqnarray}
where $k_p=g(\varepsilon/(\varepsilon+1))^{1/2}$,  $k_{pz}=ig
/\sqrt{-\varepsilon-1}$ and   $k_{pz}^m=-\varepsilon k_{pz}$. This
can, alternatively, be written as: ${\bf e }_{spp\pm}(\r)={\bf e
}_{spp\pm}^{(\nu=1)}$ for $z>0$;  and ${\bf e }_{spp\pm}(\r)={\bf e
}_{spp\pm}^{(\nu=2)}$ for $z<0$.\\
 The magnetic field associated is continuous at the interface and
equal to:
\begin{eqnarray}{\bf
h }_{spp\pm}(\r)&=&\frac{-i}{g} \nabla\times{\bf e
}_{spp\pm}(\r)\end{eqnarray}
 Now consider a lossless
metal, characterized by a  real and negative dielectric constant
$\varepsilon$ and consider a plasmon moving in the forward
direction, (the subscript + will be omitted). The incident Poyinting
vector of the plasmon in the air side is:
\begin{equation} {\ S}_{spp}^{\nu=1} =
 \int_0^\infty dz\;{\bf e }_{spp}\times{\bf h}_{spp}^* \cdot\u_x=
\; \frac{ k_p}{g}\:\frac{Z_s^2+1 }{2|k_{pz}|}
\end{equation}
while in the metal is ${ S}_{spp}^{\nu=2}=\int_{-\infty}^0 dz\:{\bf
e }_{spp}\times{\bf h}_{p}^* \cdot\u_x=Z_s^4 { S}_{spp}^{\nu=1}$
 where $ Z_s=-i/\sqrt{|\varepsilon|}$.
The total Poynting vector energy flux associated to a plasmon mode
in a lossless metal is:
\begin{eqnarray}{ S}_{spp}&=&{ S}_{spp}^{\nu=1}+{ S}_{spp}^{\nu=2}=\frac{\sqrt{-\varepsilon}}{2g}
\frac{(\varepsilon+1)(\varepsilon^2-1)}{\varepsilon^3} \geq 0.
\:\nonumber
\end{eqnarray}
\section{\label{Pmodes}P-Modes}
We shall repeat, out of completeness, the explicit expression for
p-waves, particularly in the far field when  $\k/g=\u_r$. In this
case these modes are expressed in terms of the direct space polar
angle $\alpha$ by noticing that $k_x=g \cos\alpha$ and
$k_z=k_z^{(\nu=1)}=g\sin\alpha$ in the air semi-space and
$k_z^m=k_z^{(\nu=2)}= g\sqrt{\varepsilon-\cos\alpha^2}$ in the
metal. Hence
\begin{eqnarray} {\bf
k}_p^{\pm}(\alpha)&=& \frac{1}{g} \left( k_z\u_x \mp
k_x\u_z\right)=\sin\alpha\u_x \mp \cos\alpha \u_z\nonumber
\\  \k_p^{m\pm}(\alpha)&=& \frac{1}{\sqrt{\varepsilon}g} \left( k_z^m\u_x
\mp
k_x\u_z\right)=\nonumber\\&=&\sqrt{\frac{|\varepsilon|+\cos^2\alpha}{|\varepsilon|}}\u_x
\mp \frac{\cos\alpha}{\sqrt{\varepsilon}}\u_z
\end{eqnarray}
\subsection{Reflection and Transmission coefficients for a plane
surface} For reference, we give here the Fresnel coefficients for an
air metal interface.
In the present treatment we
 only deal with the reflection coefficient for a p-wave propagating
 from air to metal, and this is :
\begin{eqnarray}r_p=r_p^{(1,1)}&=&\frac{ k_z^m- \varepsilon\,k_z}{k_z^m+ \varepsilon\,k_z}
\end{eqnarray}
where notice that, for the sake of tidiness, we omit the superscript
throughout.\\
As to the transmission coefficients the one for a wave (2,1)
propagating from the metal to air is $t_p^{(2,1)}$, while the one
for a p-wave transmitted from the air medium to the metal is
$t_p^{(1,2)}$.
\begin{eqnarray}t_p^{(2,1)}=\frac{2 k_z^m \sqrt{\varepsilon}}{k_z^m+ \varepsilon\,k_z}\;\;\;\;\;
 t_p^{(1,2)}=\frac{2 k_z \sqrt{\varepsilon}}{k_z^m+
\varepsilon\,k_z}\end{eqnarray}
 Notice that the transmission coefficients are related as follows:
\begin{eqnarray}\label{A6}\frac{ t_p^{(1,2)} }{k_z^{(1)}}=\frac{t_p^{(2,1)}}{k_z^{(2)}} \end{eqnarray}
\subsection{\label{Key}Key Identities}
 The following expressions for the reflection and transmission
 coefficients are essential to derive eq.(\ref{Gammax}) and
 eq.(\ref{Gammaz}):
\begin{eqnarray}t_p^{(1,2)}(\alpha)=\frac{2 \sqrt{\varepsilon}\sa }
{\sqrt{\varepsilon-\cos^2\alpha}+\varepsilon \sa}\end{eqnarray}

\begin{eqnarray} 1+r_p(\alpha)=\frac{2 \sqrt{\varepsilon-\cos^2\alpha} }
{\sqrt{\varepsilon-\cos^2\alpha}+\varepsilon \sin\alpha}\\[3mm]
1-r_p(\alpha)=\frac{2 \varepsilon\sin\alpha }
{\sqrt{\varepsilon-\cos^2\alpha}+\varepsilon
\sin\alpha}\end{eqnarray}

\section{\label{GT}Asymptotic Green's Tensors}
The asymptotic expressions for the Green tensor for $3$D scatterers
are found in references\cite{Andrey,NovEquivAsymp,NanoOptics}. We
have already presented the derivation scheme for bi-dimensional
defects in Appendix B of Ref.[\onlinecite{AlexImpedance}], for a
groove. As explained therein the Surface plasmon Green tensor and
the far-field Green tensor are obtained from its angular spectrum.
From the relevant Sommerfeld integral the surface plasmon
contribution is obtained by applying the residue theorem and the
far-field Green tensor instead is obtained by
applying the method of the steepest descent.\\
 For the case of the ridge we use the total Green tensor of the
 background in the vacuum semi-space. This can be written as the sum
 of the direct Green Tensor (the free space green tensor ) and the indirect
 green tensor (which gives the contribution due to the reflections at the
 metal plane interface).
 Hence \begin{eqnarray}\G^{(1)}({\bf R},\r')={\bf \hat G}_0({\bf
R},\r')+{\bf \hat G}_s({\bf R},\r')\end{eqnarray} where the spectral
representation for the direct Green tensor is:
\begin{eqnarray}{\bf \hat G}_0({\bf
R},\r')&=&\frac{i}{ 4 \pi} \int^\infty_{-\infty} \frac{dk_x}{k_z}\:
e^{i k_z( Z-z')}{e^{i k_x(X-x') }}  \: \k_p^+\k_p^+,\nonumber
\\\end{eqnarray}
while for the indirect Green Tensor:
\begin{eqnarray}{\bf \hat G}_s({\bf
R},\r')=\frac{i}{ 4 \pi} \int^\infty_{-\infty} \frac{dk_x}{k_z}\:
{e^{i k_x(X-x') }}e^{i k_z( Z+z')}  r_p \: \k_p^-\k_p^+.\nonumber
\\\end{eqnarray}
Applying the residue theorem and the steepest descent method to
$\G^{(1)}({\bf R},\r')$ we end up with eq.(\ref{asyp}) and
eq.(\ref{Gp1}) for ($\nu=1$).
\\ For the groove case we need to expand the Green Tensor connecting a
point in the metal to a point in air. This is just:
\begin{eqnarray} {\bf \hat G}^{(2)}({\bf
R},\r')=\frac{i}{ 4 \pi} \int^\infty_{-\infty}
&&\frac{dk_x}{k_z^m}\: {e^{i k_x(X-x') }}{e^{i (k_z
Z-k_z^mz')}}\nonumber\times \\ &\times& t_p^{(2,1)} \:
\k_p^{m+}\k_p^+\end{eqnarray} Applying the residue theorem and the
steepest descent method to $\G^{(2)}({\bf R},\r')$ we end up with
eq.(\ref{asyp}) and eq.(\ref{Gp1}) for ($\nu=2$) . Notice that the
form of ${{\bf \hat G}^\infty}(\alpha,{\bf r'})=-{\bf
\Upsilo}^{(\nu)} \u_\alpha $ given
in Sec.~\ref{Coeff}, is obtained by recognizing $k_p^+=-\u_\alpha$.\\
One more subtlety, that might be confusing, is how we pass from the
transmission coefficient $t_p^{(2,1)} $ in the integral to the
transmission coefficient $t_p^{(1,2)} $ in the asymptotic form ${\bf
\Upsilo}^{(2)} $ .
This comes about because when we apply the method of the steepest
descent to the integral we get eq.(\ref{asyp} ) with:
\begin{eqnarray}{{\bf \hat G}^\infty}({\bf r'}) =  \frac{k_z^m}{k_z} \,t_p^{(2,1)}\, {\bf
k}_p^m \;\u_\alpha\;=t_p^{(1,2)}\, {\bf k}_p^m \; \u_\alpha
\end{eqnarray}
where, in the last equation, we have used the identity
eq.(\ref{A6}).



\end{document}